\documentclass[sigconf]{acmart}

\usepackage[skip=1pt]{caption}
\usepackage{booktabs} 
\usepackage{ifthen}
\usepackage{color}
\usepackage{amsmath}
\usepackage{amssymb}
\usepackage{xspace}
\usepackage[T1]{fontenc}
\usepackage{enumitem}
\usepackage[linesnumbered,ruled,vlined]{algorithm2e}
 \usepackage{array,multirow,graphicx}
\usepackage{float}
\usepackage{balance}
\setcopyright{none}
\newboolean{showcomments}
\setboolean{showcomments}{true}
\ifthenelse{\boolean{showcomments}}
 { \newcommand{\mynote}[2]{
      \fbox{\bfseries\sffamily\scriptsize#1}
        {\small$\blacktriangleright$\textsf{\emph{#2}}$\blacktriangleleft$}}}
        { \newcommand{\mynote}[2]{}}

\newcommand{\toolname}{{\sc FraudDroid}\xspace}

\setcopyright{none}
\acmDOI{10.475/123_4}

\acmISBN{123-4567-24-567/08/06}

\begin{document}
\title{FraudDroid: Automated Ad Fraud Detection for Android Apps}

\author{
Feng Dong$^{1}$, Haoyu Wang$^{1 *}$, Li Li$^{2}$, Yao Guo$^{3}$, Tegawende F. Bissyande$^{4}$, \\Tianming Liu$^{1}$, Guoai Xu$^{1}$, Jacques Klein$^{4}$
}

\affiliation{%
  \institution{$^{1}$ Beijing University of Posts and Telecommunications, China
  $^{2}$ Monash University, Australia
  $^{3}$ Peking University, China
  $^{4}$ University of Luxembourg, Luxembourg\\
}
}
\thanks{* Corresponding Author}
\renewcommand{\shortauthors}{F. Dong et al.}

\begin{abstract}
Although mobile ad frauds have been widespread, state-of-the-art approaches in the literature have mainly focused on detecting the so-called \emph{static placement frauds}, where only a single UI state is involved and can be identified based on static information such as the size or location of ad views. Other types of fraud exist that involve multiple UI states and are performed dynamically while users interact with the app. Such \emph{dynamic interaction frauds}, although now widely spread in apps, have not yet been explored nor addressed in the literature.
In this work, we investigate a wide range of mobile ad frauds to provide a comprehensive taxonomy to the research community. We then propose, \toolname,  a novel hybrid approach to detect ad frauds in mobile Android apps. \toolname analyses apps dynamically to build UI state transition graphs and collects their associated runtime network traffics, which are then leveraged to check against a set of heuristic-based rules for identifying ad fraudulent behaviours.
We show empirically that \toolname detects ad frauds with a high precision ($\sim 93\%$) and recall ($\sim 92\%$). Experimental results further show that \toolname is capable of detecting ad frauds across the spectrum of fraud types. By analysing 12,000 ad-supported Android apps, \toolname identified 335 cases of fraud associated with 20 ad networks that are further confirmed to be true positive results and are shared with our fellow researchers to promote advanced ad fraud detection.
\end{abstract}

%
%

\ccsdesc[500]{Computer systems organization~Embedded systems}
\ccsdesc[300]{Computer systems organization~Redundancy}
\ccsdesc{Computer systems organization~Robotics}
\ccsdesc[100]{Networks~Network reliability}


\maketitle

\section{Introduction}
The majority of apps in Android markets are made available to users for free~\cite{measureGooglePlay, Wang-WWW-2017, appbrain-stats}. 
This has been the case since the early days of the Android platform when
almost two-thirds of all apps were free to download. Actually,
developers of third-party free apps are compensated for their work by leveraging in-app advertisements (ads) to collect revenues from ad networks~\cite{Unsafe}. The phenomenon has become common and is now part of the culture in the Android ecosystem where
advertisement libraries are used in most popular apps~\cite{measureGooglePlay}.
App developers get revenue from advertisers based either on the number of ads displayed (also referred to as {\em impressions}) or the number of ads clicked by users (also referred to as {\em clicks})~\cite{Impression}. 

While mobile advertising has served its purpose of ensuring that developers interests are fairly met, it has progressively become plagued by various types of frauds~\cite{DECAF,SmartAds}. Unscrupulous developers indeed often attempt to cheat both advertisers and users with fake or unintentional ad impressions and clicks. These are known as  {\bf ad frauds}. For example, mobile developers can typically employ individuals or bot networks to drive fake impressions and clicks so as to earn profit~\cite{Madfraud}. A recent report has estimated that mobile advertisers lost up to 1.3 billion US dollars due to ad fraud in 2015 alone~\cite{MobileAdFraud}, making research on ad fraud detection a critical endeavour for sanitizing app markets.

Research on ad frauds has been extensively carried in the realm of web applications. The relevant literature mostly focuses on \emph{click fraud} which generally consists of leveraging a single computer or botnets to drive fake or undesirable impressions and clicks. 
A number of research studies have extensively characterized click frauds~\cite{Ghost,malware12,divma11} and analysed its profit model~\cite{PharmaLeaks}. Approaches have also been proposed to detect click frauds by analysing network traffic~\cite{Detectives,SLEUTH1} or by mining search engine's query logs~\cite{SBotMiner}. 

Nevertheless, despite the specificities of mobile development and usage models, the literature on in-app ad frauds is rather limited. One example of work is the DECAF~\cite{DECAF} approach for detecting \emph{placement frauds}: these consist in manipulating visual layouts of ad views (also referred to as elements or controls) to trigger undesirable impressions in Windows Phone apps. DECAF explores the UI states (which refer to {\bf snapshots of the UI} when the app is running) in order to detect ad placement frauds implemented in the form of hidden ads, the stacking of multiple ads per page, etc. MAdFraud~\cite{Madfraud}, on the other hand, targets Android apps to detect in-app click frauds by analysing network traffic.

Unfortunately, while the community still struggles to properly address
well-known, and often trivial, cases of ad frauds, deception techniques 
used by app developers are even getting more sophisticated, as reported recently in news outlets~\cite{gfraud1,gfraud2}. Indeed, besides the aforementioned click and placement frauds, many apps implement advanced procedures for tricking users into unintentionally clicking ad views while they are interacting with the app UI elements. In this work, we refer to this type of ad frauds as \emph{\textbf{dynamic interaction frauds}}. 

\begin{figure}[t]
  \centering
  \includegraphics[width=2in]{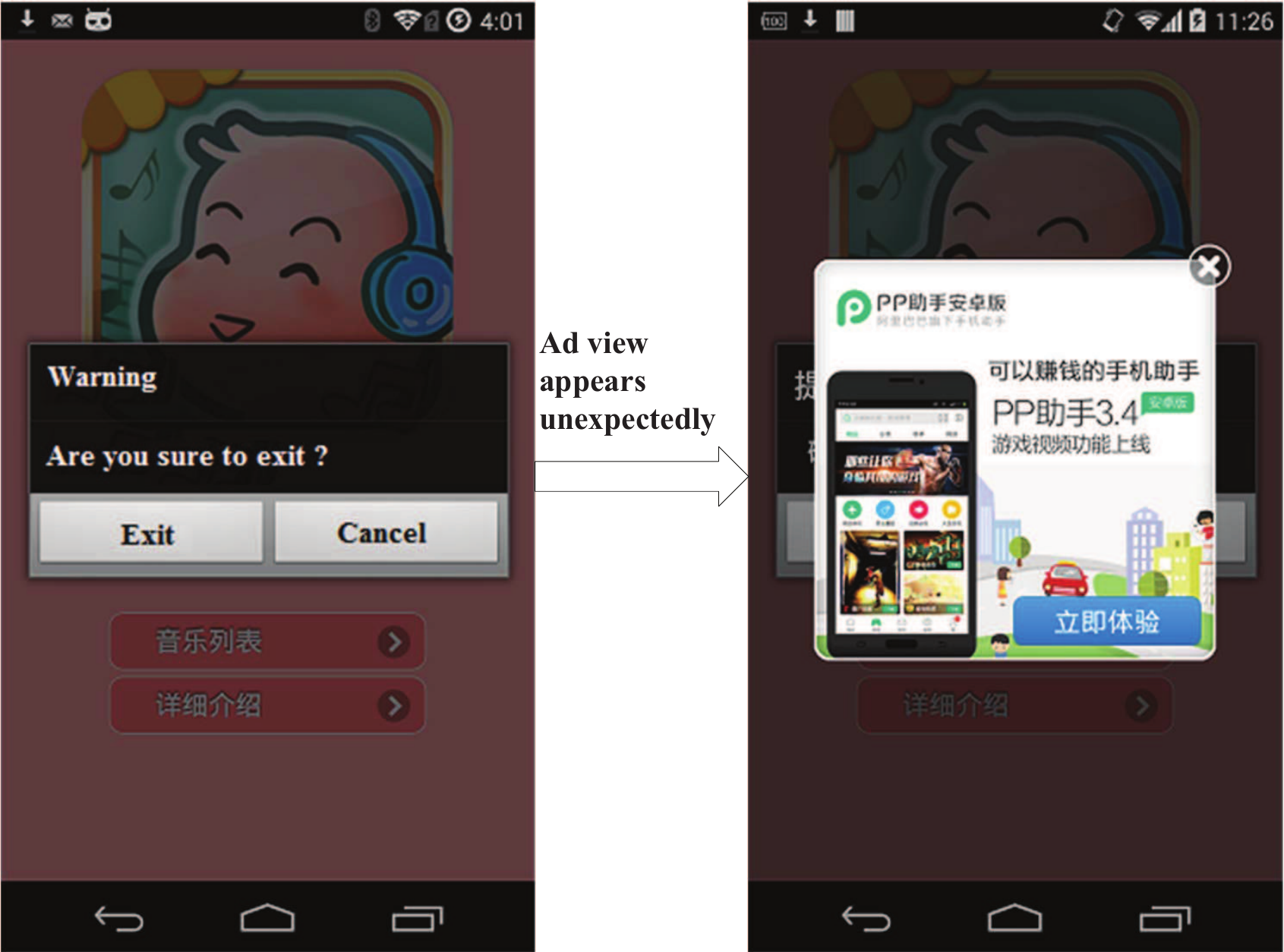}\\
  \caption{An example of interaction fraud. {\textnormal{A rapid user might accidentally click on the ad instead of the intended "Exit/Cancel".}}}
  \label{fig:interactiveExample}
\end{figure}

Figure~\ref{fig:interactiveExample} illustrates the case of the app \emph{taijiao music}\footnote{An educational music player (\emph{com.android.yatree.taijiaomusic}) targeting pregnant mothers for antenatal training.} where an ad view gets unexpectedly popped up on top of the exit button when the user wants to exit the app: this usually leads to an unintentional ad click. Actually, we performed a user study on this app and found that 9 out of 10 users were tricked into clicking the ad view.
To the best of our knowledge, such frauds have not yet been explored in the literature of mobile ad frauds, and are thus not addressed by the state-of-the-art detection approaches.


\textbf{This paper.} We perform an exploratory study of a wide range of new ad fraud types in Android apps and propose an automated approach for detecting them in market apps. To that end, we first provide a taxonomy that characterizes a variety of mobile ad frauds including both {\emph{static placement frauds}} and {\emph{dynamic interaction frauds}}. 
While detection of the former can be performed via analysing the static information of the layout in a single UI state~\cite{DECAF}, detection of the latter presents several challenges, notably for:
\begin{itemize}[leftmargin=*]
 \item \emph{{Dynamically exercising ad views in a UI state, achieving scalability, and ensuring good coverage in transitions between UI states}}: A UI state is a running page that contains several visual views/elements, also referred to as controls in Android documentation. Because dynamic interaction frauds involve sequences of UI states, a detection scheme must consider the \emph{transition between UI states}, as well as background resource consumption such as network traffic. For example, in order to detect the ad fraud case presented in Figure ~\ref{fig:interactiveExample}, one needs to analyse both current and next UI states to identify any ad view that is placed on top of buttons and which could thus entice users to click on ads unexpectedly.
Exercising apps to uncover such behaviours can however be time-consuming: previous work has shown that it takes several hours to traverse the majority UI states of an app based on existing Android automation frameworks~\cite{AMC}.

\item \emph{{Automatically distinguishing ad views among other views}}: In contrast with UI on the Windows Phone platform targeted by the state-of-the-art (e.g., DECAF~\cite{DECAF}), Android UI models are generic and thus it is challenging to identify \emph{ad views} in a given UI state since no explicit labels are provided to distinguish them from other views (e.g., text views). During app development, a view can be added to the {\tt Activity}, which represents a UI state implementation in Android, by either specifying it in the XML layout~\cite{Layouts} or embedding it in the source code. In preliminary investigations, we found that most ad views are actually directly embedded in the code, thus preventing any identification via straightforward XML analysis. 


\end{itemize}


Towards building an approach that achieves accuracy and scalability in Android ad fraud detection, we propose two key techniques aimed at addressing the aforementioned challenges:


\begin{itemize}[leftmargin=*]
  \item \emph{Transition graph-based UI exploration}. 
  This technique builds a UI transition graph by simulating interaction events associated with user manipulation. We first capture the relationship between UI states through building the transition graphs between them, then identify ad views based on call stack traces and unique features gathered through comparing the ad views and other views in UI states.
The scalability of this step is boosted by our proposed ad-first exploration strategy, which leverages probability distributions of the presence of an ad view in a UI state.

 \item \emph{Heuristics-supported ad fraud detection}. By manually investigating various real-world cases of ad frauds,
 we devise heuristic rules from the observed characteristics of fraudulent behaviour. Runtime analysis focusing on various behavioural aspects such as view size, bounds, displayed strings or network traffic, is then mapped against the rules to detect ad frauds.

\end{itemize}

These techniques are leveraged to design and implement a prototype system called \toolname for detecting ad frauds in Android apps. This paper makes the following main contributions:
\begin{enumerate}[leftmargin=*]
  \item We create a taxonomy of existing mobile ad frauds. This taxonomy, which consists of 9 types of frauds, includes not only previously studied {\emph{static placement frauds}}, but also a new category of frauds which we refer to as {\emph{dynamic interaction frauds}}.
  \item We propose \toolname, a new approach to detect mobile ad frauds based on UI transition graphs and network traffic analysis. Empirical validation on a labelled dataset of 100 apps demonstrates that \toolname achieves a detection precision of 93\% and recall of 92\%. To the best of our knowledge, \toolname is the first approach that is able to detect the five types of dynamic interaction ad frauds presented in our taxonomy.  
 \item We have applied \toolname in the wild on 12,000 apps from major app markets to demonstrate that it can indeed scale to markets. Eventually, we identified 335 apps performing ad frauds, some of them are popular apps with millions of downloads. 94 of such apps even come from the official Google Play store, which indicates that measures are not yet in place to fight fraudulent behaviour.
We have released the benchmarks and experiment results to our research community at: 
\begin{center}
\url{https://github.com/FraudDroid-mobile-ad-fraud}\footnote{We have created a new GitHub account and anonymized the author-revealing information.}
\end{center}
\end{enumerate}

\section{A Taxonomy of Mobile Ad Frauds}

\begin{figure}[!h]
  \centering
  \vspace{-0.3cm}
  \includegraphics[width=0.7\linewidth]{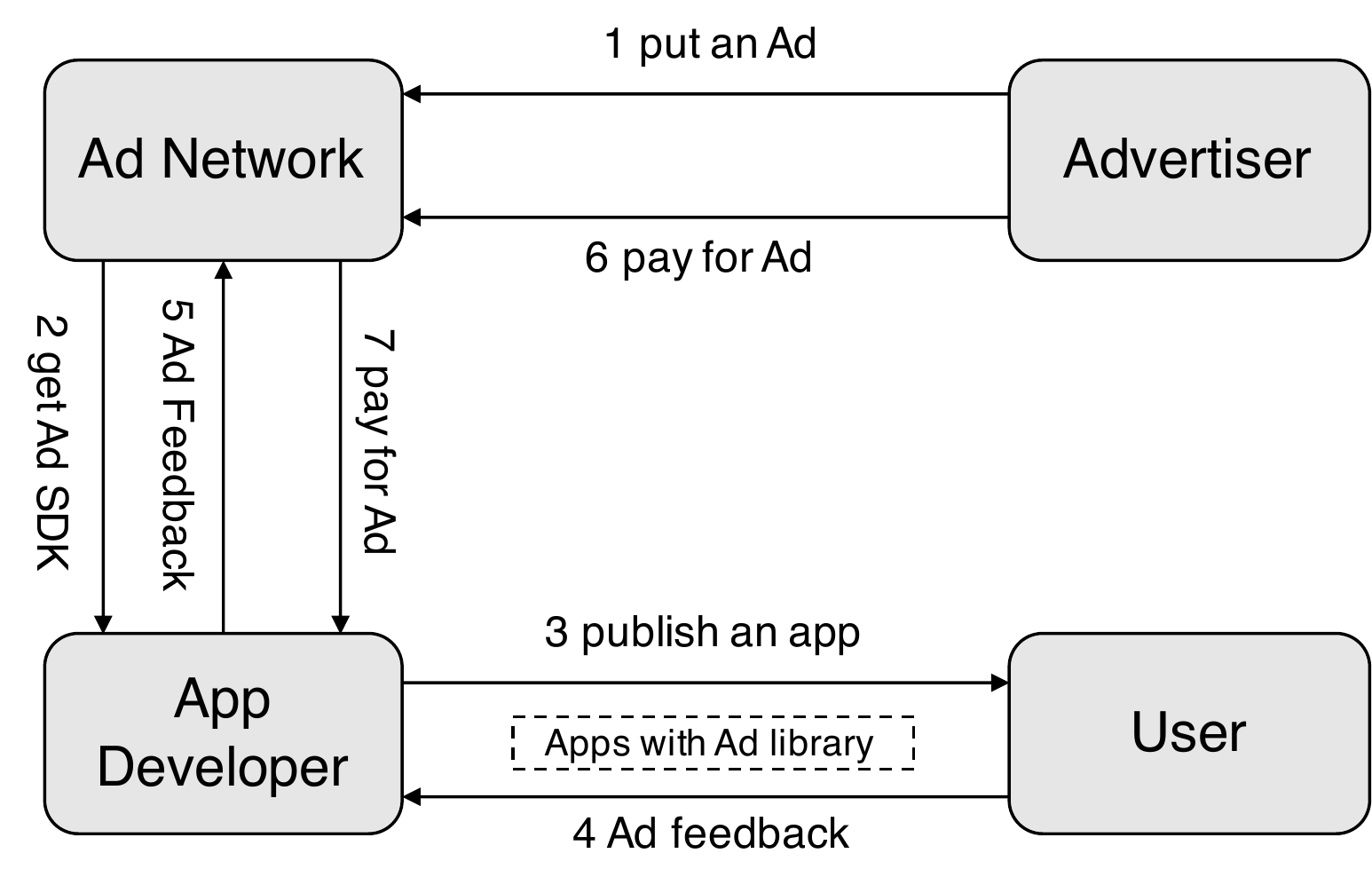}\\
  \caption{An overview of the mobile advertising ecosystem.}
  \label{fig:mobileAd}
  \vspace{-0.3cm}
\end{figure}

Before presenting the taxonomy of mobile ad frauds, we briefly overview the mobile advertising ecosystem.
Figure~\ref{fig:mobileAd} illustrates the workflow of interactions among different actors in the mobile advertising ecosystem.

\vspace{-0.1in}
\subsection{Mobile Advertising}
The role of an {\em advertiser} is to design ads that will be distributed to user devices. Such ads are displayed through third-party apps that are published by {\em app developers}. The {\em ad network} thus plays the role of a trusted intermediary platform, which connects advertisers to app developers by providing toolkits (e.g., in the form of ad libraries that fetch and display ads at runtime) to be embedded in app code. When a user views or clicks on an ad, the ad network (which is paid by advertisers) receives a feedback based on which the app developer is remunerated.


Because app developers earn revenues based on the number of ad impressions and clicks, it is tempting to engage in fraudulent behaviour to the detriment of users or/and advertisers. To avoid app developers tricking users into clicking ad views (and thus artificially increasing their revenues), ad networks have put in place some strict policies and prohibition guidelines on how ad views should be placed or used in apps. For example, violations of any of the AdMob program policies~\cite{GooglePlayDeveloper, AdMobPolicy, DongHotMobile18} is regarded as ad frauds by the Google AdMob ad network. Besides, popular app markets~\cite{GooglePlayDeveloper,HuaweiDeveloper,AliDeveloper,TencentDeveloper} have released strict developer policies on how the ads should be used in apps.
Nevertheless, unscrupulous app developers resort to new tricks to commit ad frauds that market screeners fail to detect while ad networks are un-noticeably cheated. This is unfortunate as ad frauds have become a critical concern for the experience of users, the reputation of ad networks, and the investments of advertisers.

\vspace{-0.1in}
\subsection{Ad Frauds}
\label{sec:taxonomy}
While the literature contains a large body of work on placement frauds in web applications and the Windows Phone platform, very little attention has been paid to such frauds on Android. Furthermore, dynamic interaction frauds have even not been explored to the best of our knowledge.

\begin{figure}[t]
  \centering
  \includegraphics[width=0.8\linewidth]{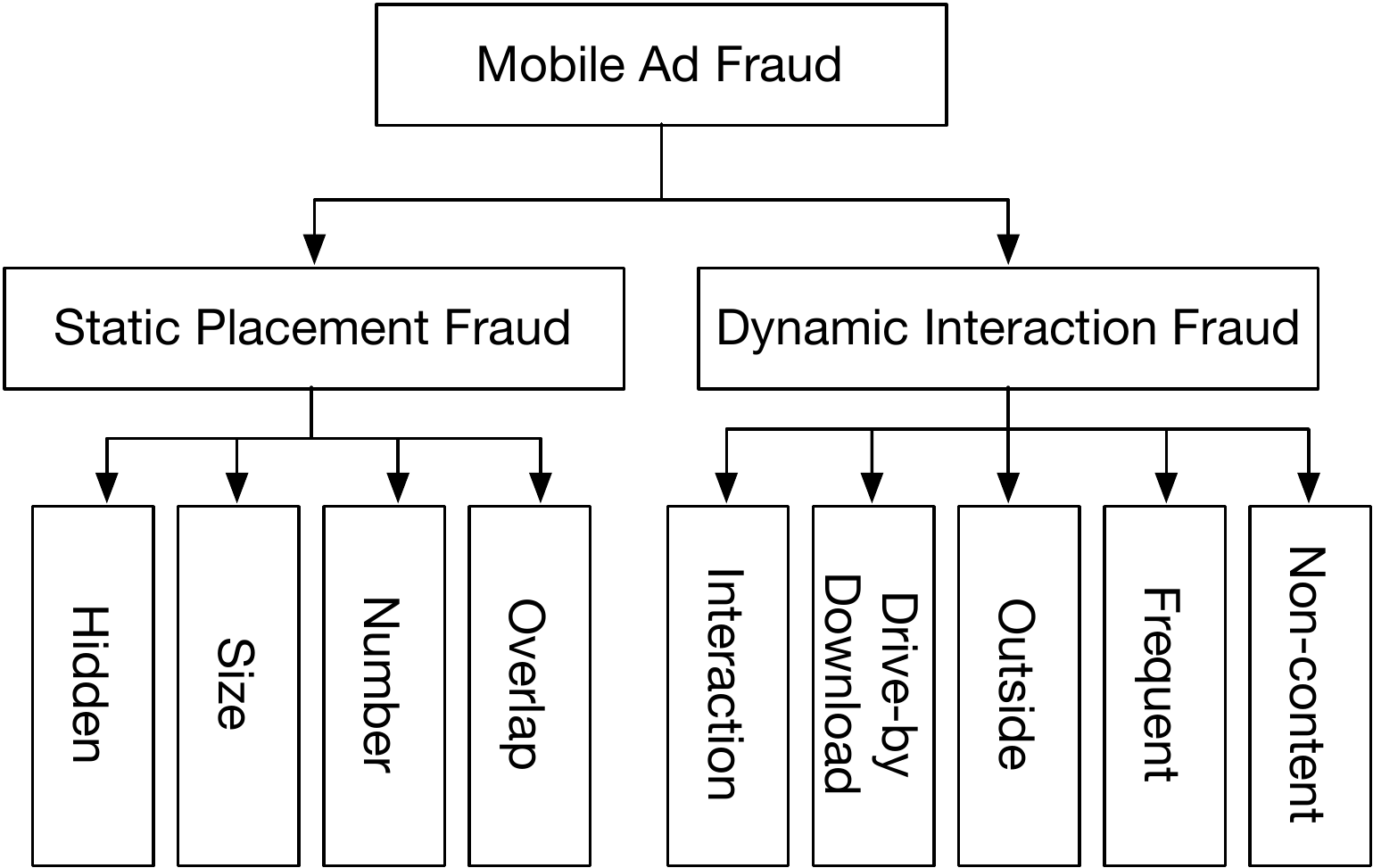}\\
  \caption{A taxonomy of mobile ad frauds.}
  \vspace{-0.2in}
  \label{fig:taxonomy}
\end{figure}

To build the taxonomy of Android ad frauds, we investigate in this work:
(1) the usage policies provided by popular ad libraries~\cite{AdMobPolicy,DoubleClick},
(2) the developer policies provided by official Google Play market~\cite{GooglePlayDeveloper} and popular third-party app markets, including Wandoujia (Alibaba App) Market~\cite{AliDeveloper}, Huawei App Market~\cite{HuaweiDeveloper} and Tencent Myapp Market~\cite{TencentDeveloper}.
(3) the guidelines on ad behaviour drafted by a communication standards association~\cite{CCSPush},
and (4) some real-world ad fraud cases.
Figure~\ref{fig:taxonomy} presents our taxonomy, which summarizes 9 different types of ad frauds, which represents by far the largest number of ad fraud types. Particularly, the five types of dynamic interaction frauds have never been investigated in the literature.

\subsubsection{Static Placement frauds}
Many fraud cases are simply performed by manipulating the ad view form and position in a UI state. ``\textbf{{\em Static}}'' implies that the detection of these frauds could be determined by static information and occur in a single UI state. ``\textbf{{\em Placement}}'' implies that the fraudulent behaviour is exploiting placement aspects, e.g., size, location, and the number of ad views, etc. We have identified four specific types of behaviour related to {\em static placement frauds}:
 \begin{enumerate}[leftmargin=*]
 \item The \textbf{\emph{Ad Hidden} fraud}. App developers may hide ads (e.g., underneath buttons) to give users the illusion of an ``ad-free app'' which would ideally provide better user experience. Such ads however are not displayed in conformance with the contract with advertisers who pay for the promotional role of ads~\cite{AdMobPolicy, AliDeveloper}.



  \item The \textbf{\emph{Ad Size} fraud}. 
  Although advice on ad size that ad networks provide is not mandatory, and there are no standards on ad size, the size ratio between the ad and the screen is required to be reasonable~\cite{AliDeveloper}, allowing the ads to be viewed normally by users~\cite{BannerAds}. Fraudulent behaviour can be implemented by stretching ad size to the limits: with extremely small ad sizes, app developers may provide the feeling of an ad-free app, however cheating advertisers; similarly, with abnormally large ad size, there is a higher probability to attract users' attention (while affecting their visual experience), or forcing them to click on the ad in an attempt to close it.
 \item The \textbf{\emph{Ad Number} fraud}. Since ads must be viewed by users as mere extras alongside the main app content, the number of ads must remain reasonable~\cite{DoubleClick, AliDeveloper}. Unfortunately, developers often include a high number of ads to increase the probability of attracting user interests, although degrading the usage experience of the app, and even severely affecting the normal functionality when ad content exceeds legitimate app content.

 \item The \textbf{\emph{Ad Overlap} fraud}. To force users into triggering undesired impressions and clicks, app developers may simply display ad views on top of actionable functionality-relevant views~\cite{AliDeveloper, AdMobPolicy, DoubleClick}. By placing ads in positions that cover areas of interest for users in typical app interactions, app developers create annoying situations where users must ``acknowledge'' the ad.
  \end{enumerate}

\begin{figure*}[t]
  \centering
  \includegraphics[width=\linewidth]{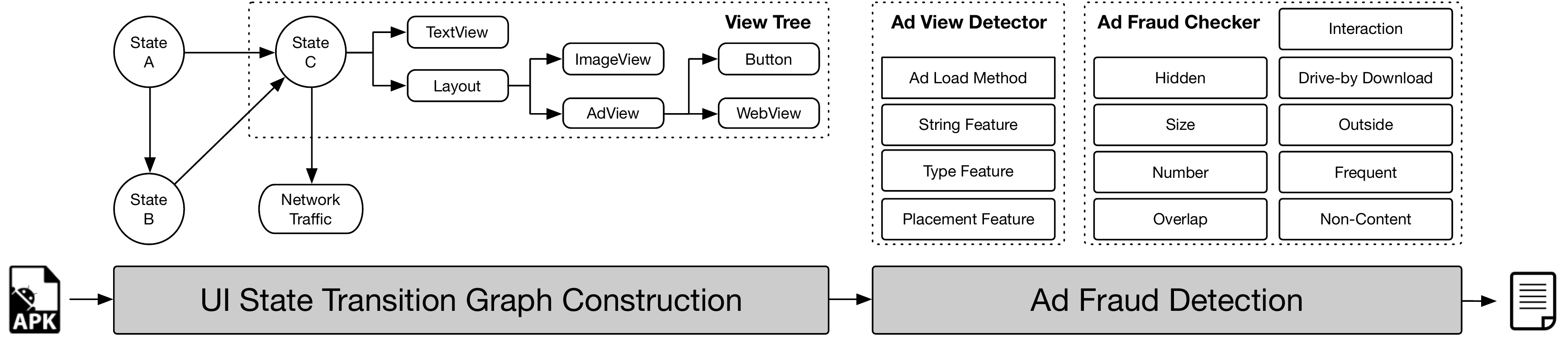}\\
  \caption{Overview of \toolname.}
  \vspace{-0.15in}
  \label{fig:overview}
\end{figure*}

\subsubsection{Dynamic Interaction Frauds}

We have also identified cases of frauds that go beyond the placement of ad views on a single UI state, but rather involve runtime behavior and may then occur in an unexpected app usage scenario. ``\textbf{{\em Dynamic}}'' implies that the detection of these frauds occur at runtime. ``\textbf{{\em Interaction}}'' implies that the fraudulent behavior is exploiting user interaction scenarios and may involve multiple UI states.
\begin{enumerate}[leftmargin=*]
\setcounter{enumi}{4}
  \item The \textbf{\emph{Interaction Ad} fraud}. In web programming, developers use interstitials (i.e., web pages displayed before or after an expected page content) to display ads. Translated into mobile programming, some ad views are placed when transitioning between UI states. However, frauds can be performed by placing interstitial ads early on app load or when exiting apps, which could trick users into accidental clicks since interaction with the app/device is highly probable at these times~\cite{AliDeveloper,AdMobPolicy,CCSPush}.

  \item The \textbf{\emph{Drive-by download Ad} fraud}. Ads are meant to provide a short promotional content designed by advertisers to attract user attention into visiting an external content page. When app developers are remunerated not by the number of clicks but according to the number of users that are eventually transformed into actual consumers of the advertised product/service, there is a temptation of fraud. A typical fraud example consists in triggering unintentional downloads (e.g., of advertised APKs) when the ad view is clicked on~\cite{CCSPush,AliDeveloper}. Such behavior often heavily impacts user experience, and in most cases, drive-by downloads cannot even be readily cancelled.

  \item The \textbf{\emph{Outside Ad} fraud}. Ads are supposedly designed to appear on pages when users use the app. Fraudulent practices exist however for popping up ads while apps are running in the background, or even outside the app environment (e.g., ad views placed on the home screen and covering app icons that users must reach to start new apps)~\cite{AliDeveloper,AdMobPolicy,CCSPush, HuaweiDeveloper}. In some extreme cases, the ads appear spuriously and the user must acknowledge them since such ads can only be closed when the user identifies and launches the app from which they come.

  \item  The \textbf{\emph{Frequent Ad} fraud}. App developers try to maximize the probability of ad impressions and clicks to collect more revenue. This probability is limited by the number of UI states in the app. Thus, developers may implement fraudulent tactics by displaying interstitial ads every time the user performs a click on the app's core content (e.g.,  even when the click is to show a menu in the same page)~\cite{AdMobPolicy, CCSPush}.

  \item The \textbf{\emph{Non-content Ad} fraud}. To maximize the number of ad impressions and trick users into unintended clicks, app developers can place ads on non-content-based pages such as thank you, error, login, or exit screens. Ads on these types of UI states can confuse a user into thinking that the ads are real app content~\cite{AdMobPolicy}.

\end{enumerate}


\section{FraudDroid}
To address ad frauds in the Android ecosystem we design and implement 
\toolname, an approach that combines dynamic analysis on UI state
 as well as network traffic data to identify fraudulent behaviours.
Figure~\ref{fig:overview} illustrates the overall architecture of \toolname. The working process unfolds in two steps: (1) {\em analysis and modelling of UI states}, and (2) {\em heuristics-based detection of ad frauds}.


To efficiently search for ad frauds, one possible step before sending apps to \toolname is to focus on such apps that have included ad libraries.
To this end, \toolname integrates a pre-processing step, which stops the analysis if the input app does not leverage any ad libraries, i.e., there will be no ad frauds in that app.
Thus we first propose to filter apps that have no permissions associated with the functioning of ad libraries, namely  {\tt INTERNET} and {\tt ACCESS\_NETWORK\_STATE}~\cite{li2016investigation}. Then, we leverage LibRadar~\cite{libradar, LibradarGithub}, a state-of-the-art, obfuscation-resilient tool to detect third-party libraries (including ad libraries) in Android apps.


%

\subsection{Analysis and Modelling of UI states}
While an app is being manipulated by users, several UI states are generated where ad views may appear. UI states indeed represent the dynamic snapshots of pages (i.e., Activity rendering) displayed on the device. One key hypothesis in \toolname is that it is possible to automate the exercise of ad views by traversing all UI states. To that end, we propose to leverage automatic input generation techniques
to collect UI information. Nevertheless, since exhaustively exercising an app is time-consuming, the analysis cannot scale to market sizes. For example, in our preliminary experiments, the analysis of a single app required several hours to complete. Empirical investigations of small datasets of ad-supported apps have revealed however that over 90\% of the UI states do not include an ad view~\cite{nath2015madscope}. We thus develop in \toolname a module, hereafter referred to as {\em FraudBot}, which implements an automated and scalable approach for triggering relevant UI states. This module exploits a fine-tuned exploration strategy to efficiently traverse UI states towards reaching most of those that contain ad views in a limited amount of time. {\em FraudBot} thus focuses on modelling a {\em UI State Transition Graph}, i.e.,  a directed graph where:

\begin{itemize}[leftmargin=*]
\item A node represents a UI state, and records information about the network traffic that this state is associated with, the trace of method calls executed while in this state, and the view tree of the UI layout in this state. 
\item An edge between two nodes represents the test input (e.g., event, view class, source state) that is associated with the transition between two states. 
\end{itemize}

\begin{figure}[!h]
  \centering
  \includegraphics[width=3.4in]{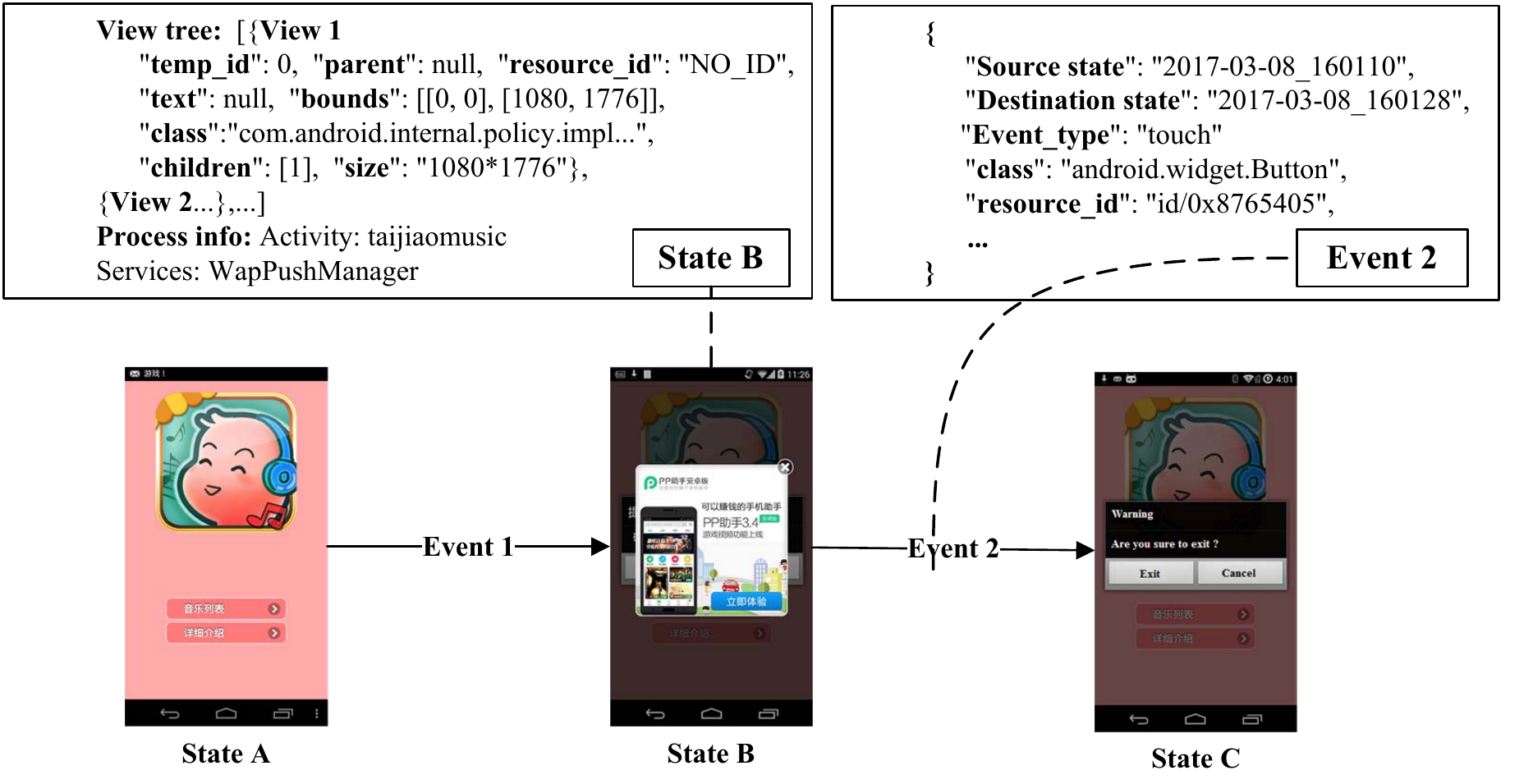}\\
  \caption{Simplified illustrative example of a UI state transition graph.}
  \vspace{-0.2in}
  \label{fig:stateTransition}
\end{figure}

Figure~\ref{fig:stateTransition} illustrates an example of UI state transition graphs, which is constructed on the fly during app automation where input event triggering state changes as well as information on the new state are iteratively added to an initially empty graph. Before discussing how \toolname exploits the UI state transition graph to unroll the app execution behaviour so as to detect ad frauds, we provide more details on the exploration strategy of \emph{FraudBot} as well as on the construction of view trees.





\subsubsection{Ad-First Exploration Strategy}
We developed a strategy that prioritizes the traversal of UI states which contain ads to account for coverage and time constraints. To simulate the behaviour of real app users, {\em FraudBot} generates test inputs corresponding to typical events (e.g., clicking, pressing) that are used to interact with common UI elements (e.g., button, scroll). 
Inspired by previous findings~\cite{nath2015madscope} that most ads are displayed in the main UI state and the exit UI state, our exploration is biased towards focusing on these states. To prioritize ad views in a state, we resort to a breadth-first search algorithm where all views in a state are reordered following ad load method traces and other ad-related features which are further described in Section~\ref{sec:adc}.
Considering that loading an ad from a remote server may take time, we set the transition time in app automation to 5 seconds, a duration that was found sufficient in our network experimental settings. By using an ad-first exploration strategy, we can reduce the app automation time from 1 hour per app to 3 minutes on average, which is a significant gain towards ensuring scalability.

\subsubsection{View Tree Construction}
\begin{figure}[!h]
  \centering
  \includegraphics[width=3.5in]{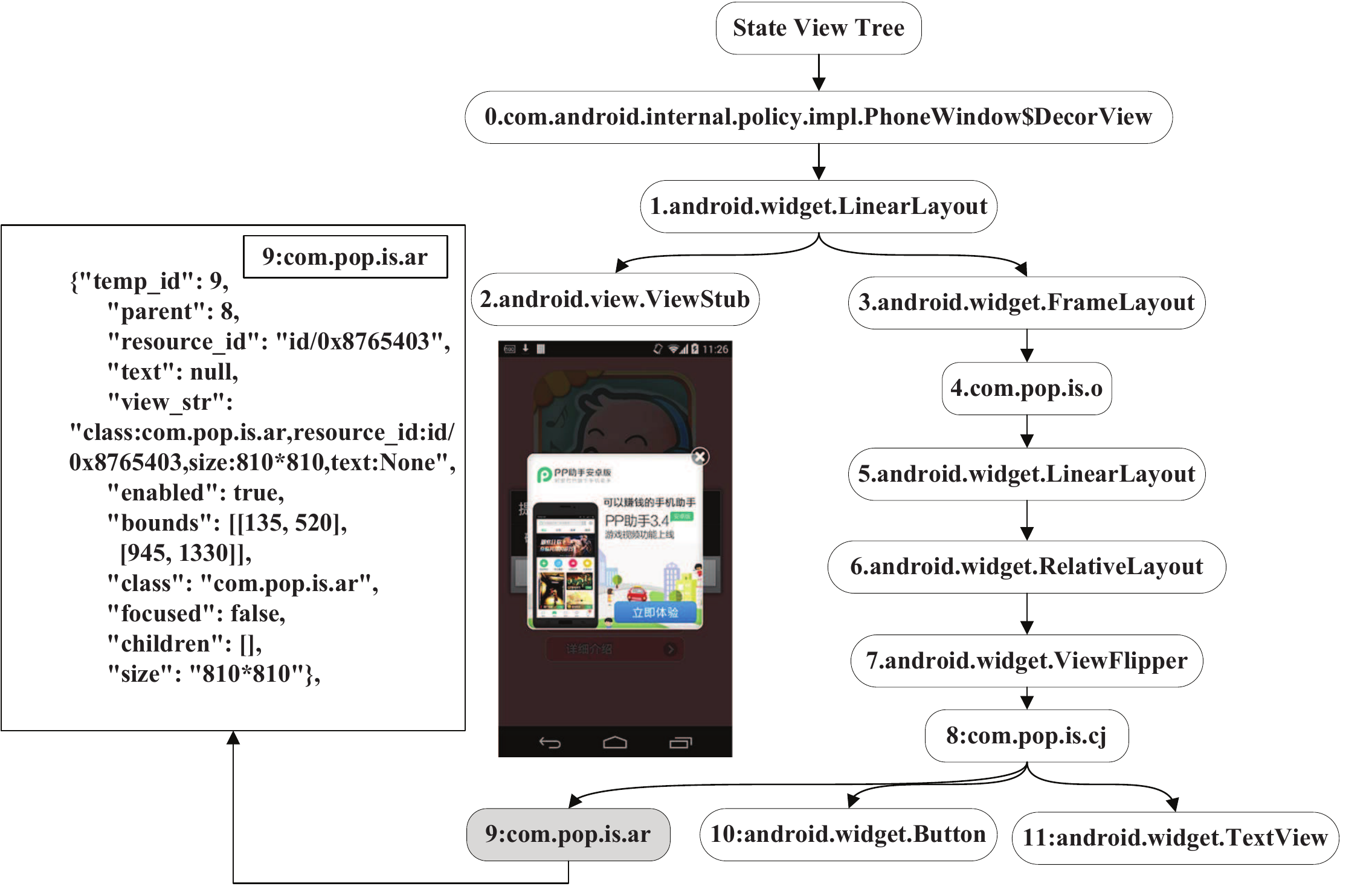}\\
  \caption{An example of view tree for interactive fraud.}
  \vspace{-0.2in}
  \label{fig:stateTransitionFraud}
\end{figure}

For each state generated by {\em FraudBot}, a view tree is constructed to represent the layout of the state aiming at providing a better means to pinpoint ad views. The tree root represents the ground layout view on top of which upper views are placed. By peeling the UI state we build a tree where each view is represented as a node: parent nodes are containers to child nodes, and users actually manipulate only leaf nodes. Each node is tagged with basic view information such as position, size, class name, etc. 
Such attributes are used to identify which nodes among the leaf nodes are likely ad view nodes. Figure~\ref{fig:stateTransitionFraud} illustrates an example of a view tree where a leaf node representing a pop-up view that is identified as an ad view.
The identification is based on the string feature representing the in-app class name (``{\tt com.pop.is.ar}'', a customized type),
the bound values of the view (\{{\tt [135, 520], [945, 1330]}\}) corresponding to the center position of the device screen, as well as the size of view ({\tt 810*810}) which is common to interstitial ads. We describe in more detail in the next section the string and placement features that are used to detect ads.


\subsubsection{Network Traffic Enhancement}
Because some ad fraudulent behaviours such as \emph{drive-by download} cannot be detected solely based on view layout information, we enhance the constructed view tree with network traffic data by connecting each view node with its associated HTTP request and the data (e.g., APK files) transmitted.
The associated network data is helpful for \toolname to detect silent downloading behaviours where certain files such as APKs are downloaded without user's interaction after an ad is clicked.

 \vspace{-0.05in}\subsection{Heuristics-based Detection of Ad Frauds}
Once a UI state transition graph is built, \toolname can find in it all necessary information for verifying whether
an ad fraud is taking place in the app. These information include, for each relevant state, the layout information, the method call trace and the network traffic information.
We implement in \toolname two modules to perform the ad fraud detection, namely \emph{AdViewDetector} for identifying ad views in a state, and \emph{FraudChecker}, for assessing whether the ad view is appearing in a fraudulent manner.


\begin{figure}[!h]
  \centering
  \includegraphics[width=1\linewidth]{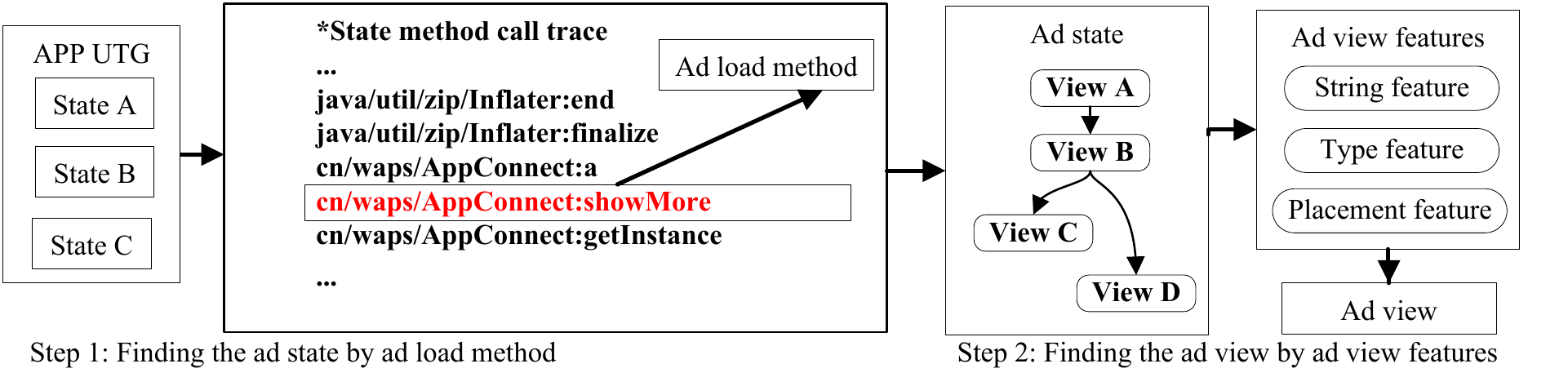}\\
  \caption{Detection of ad views \textnormal{(UTG stands for UI state transition graph)}.}
  \label{fig:adview}
  \vspace{-0.2in}
\end{figure}

\subsubsection{Detection of Ad Views}
\label{sec:adc}

We recall that Android views lack explicit labels that would allow to readily discriminate ad views from other views. 
In a previous step of \toolname, the ad-first exploration strategy was guided towards reaching UI states which are likely to include ads, as recommended by Suman Nath~\cite{nath2015madscope}.
For each of the reached UI state, the {\em AdViewDetector} module first checks if it has involved in ad-loading methods, as illustrated in Figure~\ref{fig:adview}.
After this step, most UI states are excluded from consideration as they are not really ad-contained states.
For the remaining UI states,  the {\em AdViewDetector} module of \toolname again excludes irrelevant states that are not ad-contained ones, where all the leaf nodes in the view tree associated to the UI states are checked against common ad features (i.e., String, Type and Placement features).


We have identified relevant ad features after manually labelling a large amount of ad views and normal views that we have compared them from different aspects (namely string, type and placement features). Table~\ref{tab:adfeature} presents some sample values for the ad detection features considered. 

\begin{table*}[t]
  \caption{Ad features.}
  \label{tab:adfeature}
  \begin{tabular*}{450pt}{llp{11cm}}
    \toprule
    \textbf{Aspects} & \textbf{Attribute}& \textbf{Value} \\
    \midrule
    String & Resource\_id& AdWebview, AdLayout, ad\_container, fullscreenAdView, FullscreenAd, AdActivity, AppWallActivity, etc.\\
    \hline
    Type&  Class &  ImageView, WebView, ViewFlipper\\
    \hline
    placement& Size (Bounds) & 620*800[Center], 320*50[Top or Bottom], 1776*1080[Full screen], etc.\\
    \bottomrule
  \end{tabular*}
\vspace{-0.2in}
\end{table*}

\begin{itemize}[leftmargin=*]
\item {\bf String}. Manual investigations of our labelled datasets revealed that String attributes in view definitions, such as associated class name, can be indicative of whether it is an ad view. Most ad libraries indeed often implement specialized classes for ad views instead of directly using Android system views (\emph{``android.widget.Button''}, \emph{``android.widget.TextView''}, etc.). We noted however that these specialized classes' names reflected their purpose with the keyword ``ad'' included: e.g., \emph{``AdWebview''}, \emph{``AdLayout''}. Since simply matching the keyword ``ad'' in class names would lead to many false positives with common other words (e.g., shadow, gadget, load, adapter, adobe, etc.), we rely on a whitelist of English words containing ``ad'' based on the Spell Checker Oriented Word Lists (SCOWL)~\cite{SCOWL}. When a view is associated to a class name matches the string hint ``ad'' but is not part of the whitelist, we take it as a potential ad view. 
\item {\bf Type}. Since some ad views may also be implemented based on Android system views, we investigate which system views are most leveraged. All ad views in our labelled dataset are associated to three types: ``ImageView''~\cite{imageview}, ``WebView''~\cite{webview} and ``ViewFlipper''~\cite{viewflipper}, in contrast to normal views which use a more diverse set of views. We therefore rely on the type of view used as an indicator of probable ad view implementation.
\item {\bf Placement}. In general, ad views have special size and location characteristics, which we refer to as placement features. Indeed, mobile ads are displayed in three common ways: 1) {\em Banner ad} is located in the top or bottom of the screen; 2) {\em Interstitial ad} is square and located in the centre of the screen; 3) {\em Full-screen ad} fills the whole screen. Furthermore, ad networks actually hard-code in their libraries the size for the different types of views implemented\cite{AdMobPolicy}, which are also leveraged in \toolname to identify possible ad views.
\end{itemize}

Using the aforementioned features, we propose a heuristic-based approach to detect ad views from the view tree extracted from a UI state. 
First, string and type features help to identify respectively customised views that are good candidates for ad views. Subsequently, placement features are used to decide whether a candidate ad view will be considered as such by \toolname. 
We have empirically confirmed that this process accurately identifies ad views implemented by more than 20 popular ad libraries, including Google Admob and Waps. 

\subsubsection{Identification of Fraudulent Behaviour}
Once ad views are identified across all UI states in the UI State Transition Graph, the {\em FraudChecker} module can assess their tag information to check whether a fraudulent behaviour can be spotted. This module of \toolname is designed to account for the whole spectrum of frauds enumerated in the taxonomy of Android ad frauds (cf. Section~\ref{sec:taxonomy}). To that end, {\em FraudChecker} implements heuristics rules for each specific ad fraud type:

\emph{Ad Hidden.} To detect such frauds, {\em FraudChecker} iteratively checks whether any ad view has boundary coordinate information which would indicate that it is (partially or totally) covered by any other non-ad views: i.e., it has a z-coordinate below a non-ad view and the 2-D space layout occupied by both views intersect at some point.
 
    


\emph{Ad Size.} 
Although the ad view size may vary on different devices, our manual investigation has shown that the size ratio between ad view and the screen is relatively stable for legitimate ads. We empirically define a ratio of [0.004, 0.005] for banner ads, [0.2, 0.8] for interstitial ads and [0.9, 1] for full-screen ads.
{\em FraudChecker} uses these values as thresholds, combined with position information (cf. Taxonomy in Section~\ref{sec:taxonomy}), to determine whether there is a fraudulent behaviour. Note that these values are configurable.

\emph{Ad Number.} In a straightforward manner, {\em FraudChecker} verifies that a basic policy is respected: the combined space occupied by all ads in a UI state must not exceed the space reserved to app content. When there are several ad views in a UI state along side app content (i.e., all as view tree leafs), the size of ad views in total should not exceed 50\% of the screen size.

\emph{Ad Overlap.} Similarly to the case of {Ad Hidden}, {\em FraudChecker} checks whether the space occupied by an ad view on the screen intersects with the space of other normal views. In contrast to {\em Ad Hidden}, the overlapping views are on the same level in z-coordinate. 

\emph{Interactive Ad.} This fraud occurs with an interstitial ad. {\em FraudChecker} first traverses the UI state transition graph to identify any UI state that
contains interactive views (e.g., dialogue or button) which are tempting for users to click. 
Then, it checks the next UI state in the transition graph to inspect whether there is an ad view that is placed on top of the aforementioned interactive views (i.e., dialogue or button from the preceding UI state).
If so, \toolname flags this behaviour as an interactive fraud.

\emph{Drive-by download Ad.} This fraud consists in triggering unwanted downloads (of apps or other files) without any confirmation by users after an ad is clicked on. {\em FraudChecker} flags a given UI state as \emph{drive-by download} as long as the following conditions are met: 1) there are ad views in this state; 2) there is a downloading behaviour; 3) the next state in the transition graph is still associated with the same Activity (e.g., it does not switch to other interfaces); and 4) the state is triggered by a touch event. 

\emph{Outside Ad.} To detect such frauds, {\em FraudChecker} keeps track of all Activity names that are associated with the app. By going through all UI states, it checks whether a UI state contains an ad view that has its associated Activity name different from any of the expected ones. This indicates that the ad is shown outside the app environment.

\emph{Frequent Ad.} We consider an app is suspicious to frequent ad fraud as long as it has interstitial ads or full-screen ads displayed more than three times\footnote{This number is configurable to \toolname.} in the UI state transition graph, where the three displays are triggered by different UI state transitions (i.e., several visits via the same transition path to the same ad-contained UI state are considered as one visit).

\emph{Non-content Ad.} \emph{FraudChecker} flags such UI states as \emph{Non-content Ad fraud} when it identifies that interstitial ads or full screen ads exist before or after launch/login/exit states.

\subsection{Implementation}
We have implemented a lightweight UI-guided test input generator to dynamically explore Android apps with a special focus on UI states. The UI-guided events are generated according to the position and type of UI elements. Regarding network traffic, we leverage \emph{Tcpdump}~\cite{Tcpdump} and \emph{The Bro Network Security Monitor}~\cite{Bro}, respectively, to harvest and analyse the network traffic.
\section{Evaluation}
We evaluate the effectiveness of \toolname with respect to its capability to accurately detect ad frauds (cf. Section~\ref{sec:accuracy}), and its scalability performance (cf. Section~\ref{sec:performance}) through measurements on runtime and memory cost of running \toolname on a range of real-world apps. Eventually, we run \toolname in the wild, on a collected set of 12,000 ad-supported apps leveraging 20 ad networks, in order to characterize the spread of ad frauds in app markets (cf. Section~\ref{sec:scalability}).
All experiments are performed on a real physical device, namely a Nexus 5 smartphone.
We do not use emulators since ad libraries embed checking code to prevent ad networks from serving ads when the app is being experimented on emulator environments~\cite{evading}.



\vspace{-0.1in}
\subsection{Detection Accuracy}
\label{sec:accuracy}
\begin{figure}[!h]
  \centering
  \includegraphics[width=0.8\linewidth]{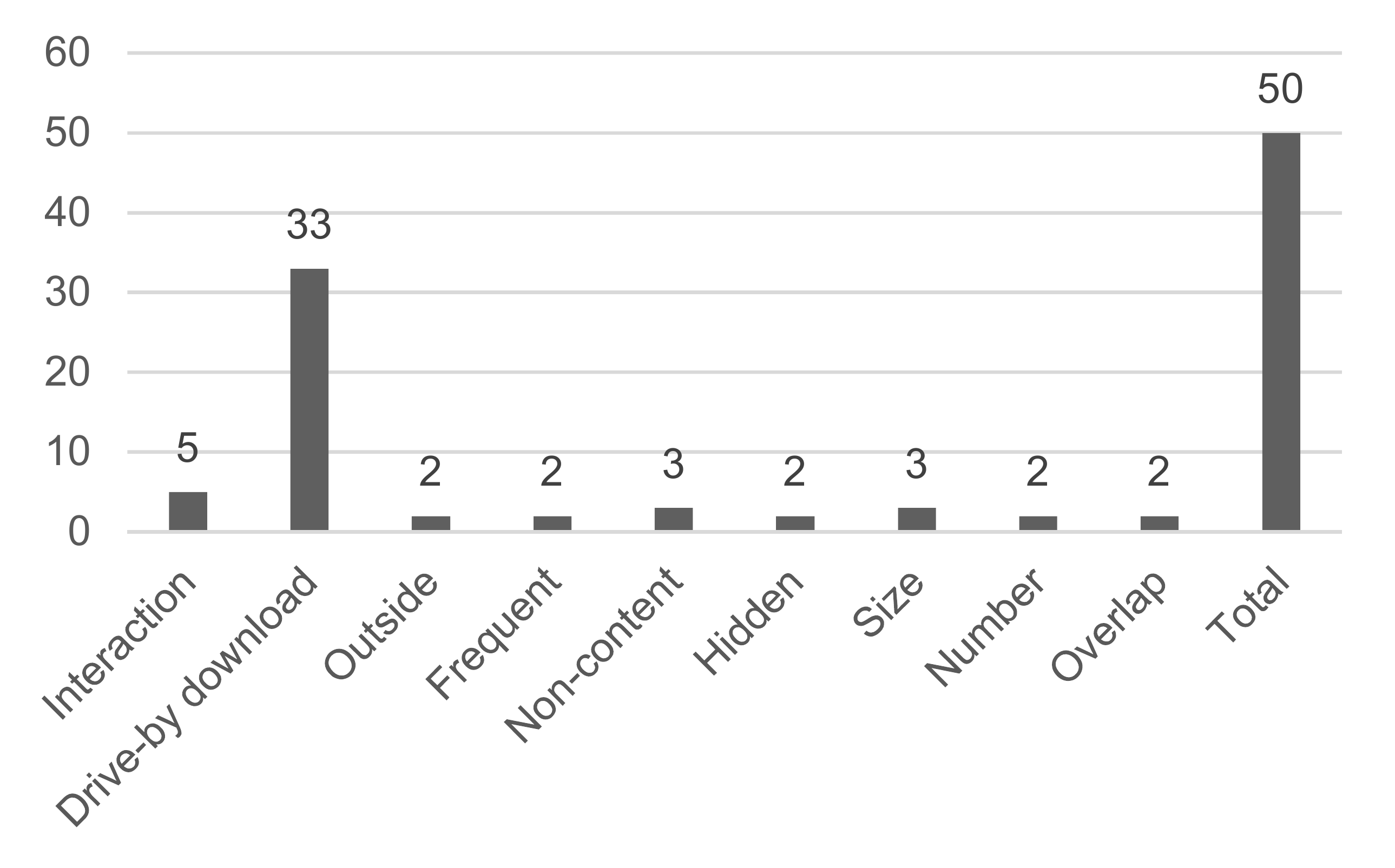}\\
  \caption{Distribution of the 50 labelled apps via their ad fraud types.} 
  \vspace{-0.2in}
  \label{fig:dataDistri}
\end{figure}

Accuracy is assessed for both the detection of ad views by {\em AdViewDectector} and the actual identification of frauds by {\em FraudChecker}. In the absence of established benchmarks in this research direction, we propose to manually collect and analyse apps towards building benchmarks that we make available to the community:
\begin{enumerate}
	\item Our first benchmark set, \emph{AdViewBench}, includes 500 apps that were manually exercised to label 4,403 views among which 211 are ad views.
	\item Our second benchmark set, \emph{AdFraudBench}, includes 100 ad-supported apps, 50 of which exhibit fraudulent behaviours that are manually picked and confirmed. To select the benchmark apps, we have uploaded more than 3,000 apps that use ad libraries to VirusTotal~\cite{VirusTotal}, and manually checked the apps that are labelled as \emph{AdWare} by at least two engines from VirusTotal. We selected apps to cover all 9 types of frauds, from the static placement and dynamic interactive categories, and ensure that each type of fraud has at least two samples. Figure~\ref{fig:dataDistri} illustrates the distribution of apps across the fraud types. We have obtained an overall of 54 ad fraud instances for 50 apps: some apps in the benchmark set indeed perform more than one types of ad frauds\footnote{For example, app
\texttt{az360.gba.jqrgsseed} is found to implement both an interaction and a drive-by download fraud.}. 
\end{enumerate}




\subsubsection{Accuracy in ad view identification}
We run the {\em AdViewDetector} module on the UI states associated to the {\em AdViewBench} of 4,403 views. The detector reports 213 views as ad views, 197 of which are indeed ad views according to the benchmark labels. The detector however misses to identify 11 ad views in the benchmark. Table~\ref{table:resultAdView} summarizes the results on precision and recall of {\em AdViewDetector}. With over 90\% of precision and recall, \toolname is accurate in detecting ad views, an essential step towards efficiently detecting ad frauds.


\begin{table}[!h]
\newcommand{\tabincell}[2]{\begin{tabular}{@{}#1@{}}#2\end{tabular}}
\centering
\caption{Confusion matrix of AdViewDetector on the {\em AdViewBench} benchmark.}
\label{table:resultAdView}
\resizebox{\linewidth}{!}{
\begin{tabular}{ |c r |c c|}
  \cline{3-4}
 \multicolumn{1}{c}{} & & \multicolumn{2}{c|}{\bf Predicted views} \\
  \cline{3-4}
 \multicolumn{1}{c}{}& & Ad-contained & Ad-free\\
\hline
\multirow{3}{*}{\rotatebox[origin=c]{90}{\parbox[c]{1.2cm}{\centering \bf Labelled views}}}  & \multicolumn{1}{|c|}{Ad-contained} & 197 True Positives & 11 False Negatives  \\
& \multicolumn{1}{|c|}{}&  & \\
& \multicolumn{1}{|c|}{Ad-free} & 14 False Positives & 4181 True Negatives \\
\hline
\end{tabular}}
\vspace{-0.15in}
\end{table}


We further investigate the false positives and false negatives results by the {\em AdViewDetector} module. On the one hand, we found by analysing the UI states collected during app automation that some ads are not successfully displayed: the library fails to load ad content due to network connection failures or loading time-outs.
Nevertheless, because of the calls to ad loading methods, \toolname will anyway flag those views (without ads displayed) as ad views, which however are not considered as such during our manual benchmark building process, resulting in false positive results. 
On the other hand, we have found that some UI states, although they include ad views, do not include calls to ad loading methods: the ad views were inherited from prior states, and some events (e.g., such as drag or scroll events) that triggered the transition to new states do not cause a reloading of views. Such cases result in false negatives by {\em AdViewDetector}, which misses to report any ad view in the UI state.




\subsubsection{Accuracy in Ad fraud detection}
To evaluate the accuracy of our approach in detecting ad fraud behaviours, we run \toolname on the \emph{AdFraudBench} benchmark apps. Table~\ref{table:resultFraudChecker} provides the confusion matrix obtained from our experiments in classifying whether an app is involved in any ad fraud. Among the 100 apps in the benchmark, \toolname flags 49 as performing ad frauds: checking against the benchmarks, these detection results include 3 cases of false positives and 4 false negatives, leading to precision and recall metrics of 93.88\% and 92\% respectively.

\begin{table}[!h]
\newcommand{\tabincell}[2]{\begin{tabular}{@{}#1@{}}#2\end{tabular}}
\centering
\caption{Confusion matrix of fraud detection by \toolname on the {\em AdFraudBench} benchmark.}
\label{table:resultFraudChecker}
\resizebox{\linewidth}{!}{
\begin{tabular}{ |c r |c c|}
  \cline{3-4}
 \multicolumn{1}{c}{} & & \multicolumn{2}{c|}{\bf Predicted frauds} \\
  \cline{3-4}
 \multicolumn{1}{c}{}& & Ad fraud & w/o Ad fraud\\
\hline
\multirow{3}{*}{\rotatebox[origin=c]{90}{\parbox[c]{1.2cm}{\centering \bf Labelled frauds}}}  & \multicolumn{1}{|c|}{Ad fraud} & 46 True Positives & 4 False Negatives  \\
& \multicolumn{1}{|c|}{}&  & \\
& \multicolumn{1}{|c|}{w/o Ad fraud} & 3 False Positives & 47 True Negatives \\
\hline
\end{tabular}}
\vspace{-0.1in}
\end{table}

We further investigate the false positive/negative detection cases to understand the root causes of \toolname's failures. We find that all such cases are caused by misclassification results by the {\em AdViewDetector} module.
\toolname misses ad frauds because the UI state is not marked as relevant since no ad views were identified on it. Similarly, when \toolname falsely identified an ad fraud, it was based on behaviours related to a view that was wrongly tagged as an ad view.

\subsection{Scalability Assessment}
\label{sec:performance}
\begin{figure}[!h]
  \centering
  \includegraphics[width=3.5in]{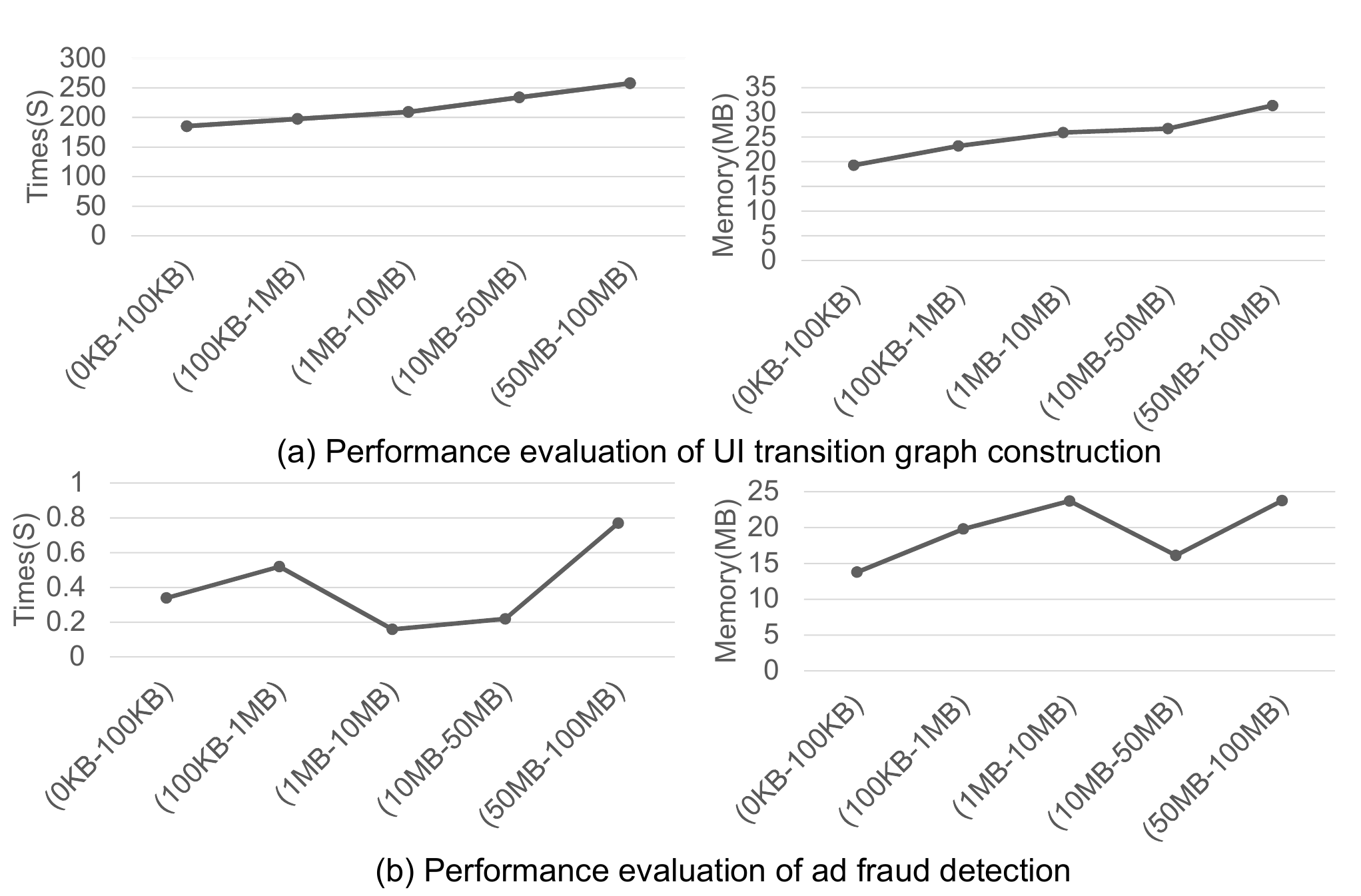}\\
  \caption{Performance evaluation of \toolname.}
  \vspace{-0.25in}
  \label{fig:performance}
\end{figure}

To assess the capability of \toolname to scale to app markets, we measure
the runtime performance of the steps implementing the two key techniques in our approach: (1) the UI state transition graph construction, and (2) the heuristics-based detection of ad frauds.
To comprehensively evaluate time and memory consumed by \toolname, we consider  apps from a wide size range. We randomly select 500 apps in each of the following five size scales\footnote{E.g., for the scale 100KB, the apps must be of size between 80KB and 120KB.}: 100KB, 1MB, 10MB, 50MB and 100MB. Overall, the experiments are performed on 2,500 apps for which we record the time spent to complete each step and the memory allocated. The results are presented in Figure~\ref{fig:performance}. 

The average time cost for constructing UI transition graphs ($216.7$ seconds) is significantly higher than that of detecting ad frauds ($0.4$ seconds). However, we note that, unlike for UI transition graphs, time to detect ad frauds is not linearly correlated to the app size. The UI transition graph is built by dynamically exploring the app where larger apps usually have more running states while detection solely relies on a pre-constructed graph.
Nevertheless, thanks to the ad-first exploration strategy, \toolname is able to analyse every apps in the dataset within 3.6 minutes. Comparing with experiments performed in state-of-the-art app automation works~\cite{Yet,DECAF}, \toolname provides substantial improvements to reach acceptable performance for scaling to app market sizes. Interestingly, memory consumption is reasonable (around 20MB), and roughly the same for both steps.

\subsection{Detection In-the-Wild}
\label{sec:scalability}

We consider over 3 million Android apps crawled between April 2017 and August 2017 from eight app markets, namely the official Google Play store and alternative markets  by Huawei, Xiaomi, Baidy, Tencent, PP, Mgyapp and 360 Mobile. 
Using LibRadar~\cite{libradar}, we were able to identify ad libraries from 20 ad networks represented each in at least 500 apps.
For the experiments in the wild, we randomly select 500 apps per ad network, except for Admob, for which we consider 2,500 apps to reflect the substantially larger proportion of ad-supported apps relying on Admob. Eventually, our dataset is formed by 12,000 Android apps, as shown in Table~\ref{table:dataset}.




\subsubsection{Overall Results.} 
{\toolname} identified 335 (2.79\%) apps among the 12,000 dataset apps as implementing ad frauds. We note that ad frauds occur on a wide range of app categories, from Games to Tools. Some detected ad-fraud apps are even popular among users with over 5 million downloads. For example, the app \emph{Ancient Tomb Run}~\cite{fraudAppTempleqqq} has received over 5 million downloads, although \toolname has found that it contains a non-content ad fraud.  

\subsubsection{Ad Fraud Distribution by Type.}
\begin{figure}[!h]
  \centering
  \includegraphics[width=\linewidth]{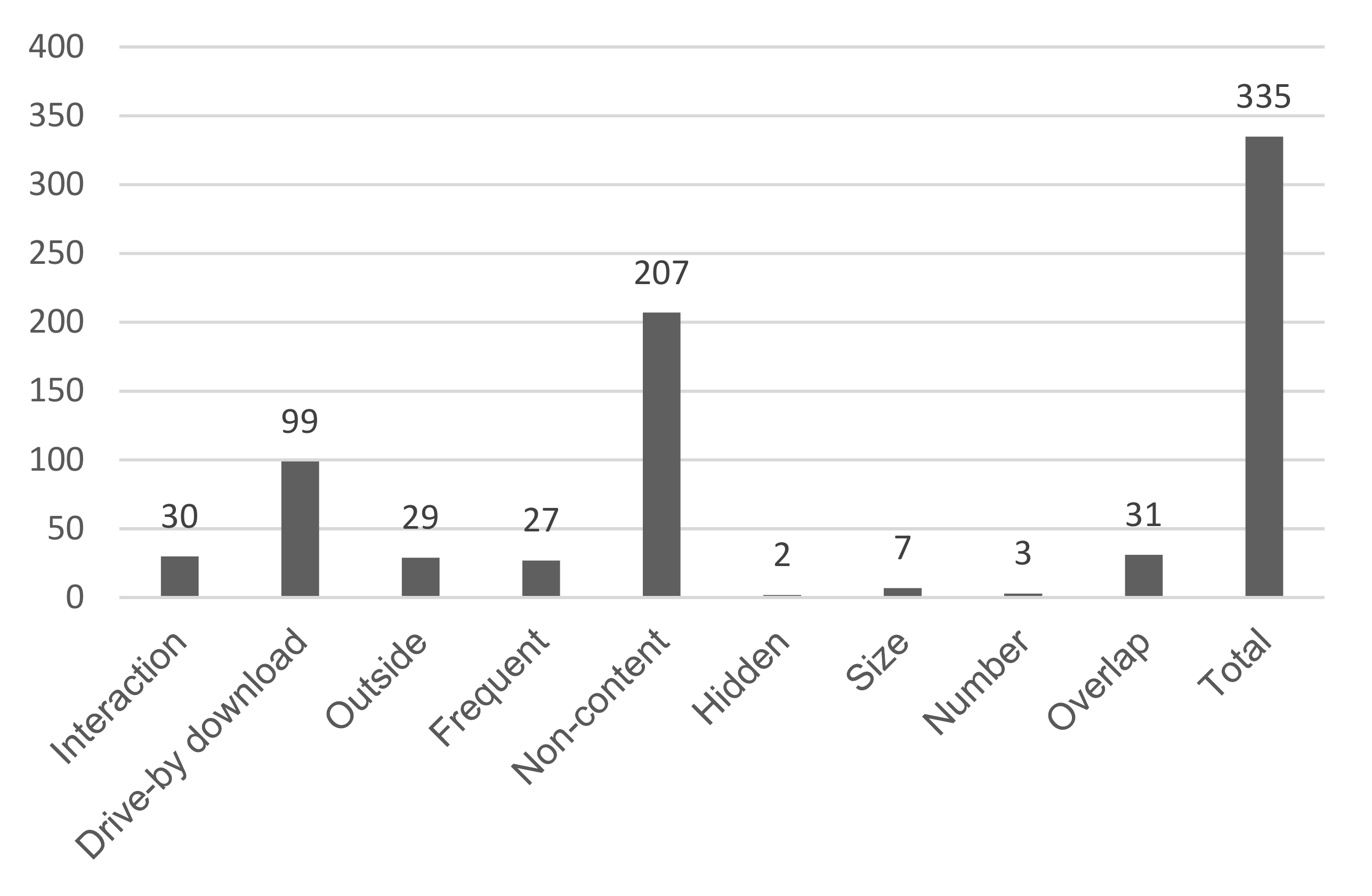}\\
  \caption{Ad fraud distribution by type.}
  \vspace{-0.15in}
  \label{fig:fraudtype}
\end{figure}

Figure~\ref{fig:fraudtype} shows the distribution of the apps based on the types of frauds. Static placement frauds only account for a small portion of the total detected ad fraud apps. Over 90\% of the fraud cases are \emph{dynamic interaction frauds}, which, to the best of our knowledge, we are the first to investigate with this work.
This is an expected result, since dynamic interaction frauds (1) are more difficult to detect, and thus they can easily pass vetting schemes on app markets, and (2) they are most effective in misleading users into triggering ad impressions and clicks, leading to more revenues.

\subsubsection{Ad Fraud Distribution by App Markets.}
Table~\ref{table:dataset} enumerates the distribution of detected ad-fraud apps based on the markets where they were collected. The \textit{PP Assistant} market includes the highest rate (4.84\%) of ad frauds among its ad-supported apps in our study. Huawei Market includes the lowest rate (0.87\%) of ad frauds. Finally the ad fraud rate in Google Play is slightly lower than the average across markets.


\begin{table}[!h]
 \newcommand{\tabincell}[2]{\begin{tabular}{@{}#1@{}}#2\end{tabular}}
 \centering
 \caption{Dataset from eight major Android app markets. 
 }
 \label{table:dataset}
 \begin{tabular}{ r c c c}
 \hline
 App Market & \#Ad Apps &\#Fraud & Percentage\\
 \hline
 Google Play & 3,630 & 94 & 2.59\%\\
 360 Mobile Assistant & 1,012 & 25 & 2.47\%\\
 Baidu Mobile Assistant  & 1,588 & 55 & 3.46\% \\
 Huawei  & 1,145 & 10 & 0.87\% \\
 Xiaomi  & 1,295 &29 & 2.24\% \\
 Tencent Yingyongbao & 843 & 18 & 2.14\% \\
 Mgyapp &1,288 &46 & 3.57\% \\
 PP Assistant & 1,199 &58 & 4.84\% \\
 \hline
 \hline
 \textbf{\emph{Total}} & \textbf{\emph{12,000}} & \textbf{\emph{335}} & \textbf{\emph{2.79\%}} \\
 \hline
 \end{tabular}
 \vspace{-0.15in}
\end{table}

These statistics show that no market is exempt from ad frauds, suggesting that, at the moment, markets have not implemented proper measures to prevent penetration of fraudulent apps into their databases. We also found that some markets do not even provide explicit policies to regulate the use of ads in apps. 

\begin{figure*}[t]
  \centering
  \includegraphics[width=0.8\textwidth]{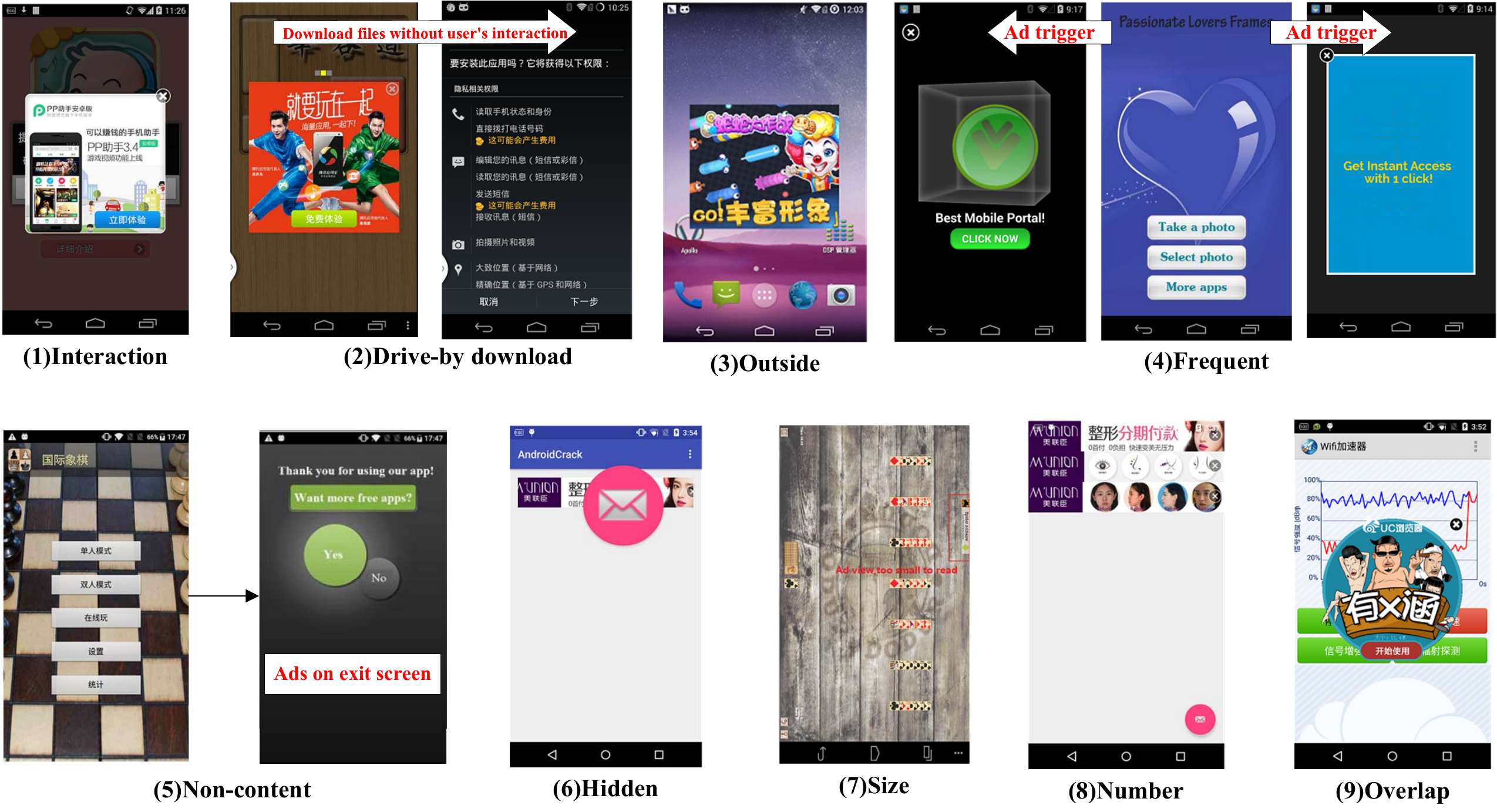}\\
  \caption{Case studies of ad-fraud apps.}
  \vspace{-0.15in}
  \label{fig:casestudy}
\end{figure*}

\subsubsection{Ad fraud Distribution by Ad Networks.}
Table~\ref{table:AdNetwork} presents the distribution of ad-fraud apps by ad networks. Although most ad frauds target some popular ad networks, no ad networks were exempt from fraudulent behaviours: \emph{Appbrain} appears to be the most targeted with 13.6\% of its associated apps involved in ad frauds. 

\begin{table}[!h]
\newcommand{\tabincell}[2]{\begin{tabular}{@{}#1@{}}#2\end{tabular}}
\centering
\vspace{-0.1in}
\caption{Distribution of ad-fraud apps based on ad networks.}
\label{table:AdNetwork}
\resizebox{\linewidth}{!}{

\begin{tabular}{ r c c | r c c}
\hline
Ad network & \tabincell{l}{\#Ad\\fraud}& \%Percent & Ad network & \tabincell{l}{\#Ad\\fraud}& \%Percent\\
\hline
Admob & 113  &4.52\%& Adwhirl & 2&0.4\%\\
Appbrain & 68 &13.6\%& Dianjin & 9&1.8\%\\
Waps & 48 &9.6\%& Vpon & 1&0.2\%\\
feiwo & 29 &5.8\%& Inmobi & 6&1.2\%\\
BaiduAd & 10 &2\% & Apperhand & 8&1.6\%\\
Anzhi & 7&1.4\% & Startapp &5&1\% \\
Youmi & 8 &1.6\%& Mobwin &2&0.4\%\\
Doodlemobile & 4&0.8\% & Jumptap &1&0.2\%\\
Adsmogo & 5 &1\%& Fyber  & 1&0.2\%\\
Kugo & 7 &1.4\%& Domob & 1&0.2\%\\
\hline
\end{tabular}
}
\vspace{-0.2in}
\end{table}



Interestingly, although \texttt{Google Admob} has published strict policies~\cite{AdMobPolicy} on how ad views should be placed to avoid ad frauds, we still found 113 fraudulent apps associated to \texttt{Admob}.
In some cases, we have found that several popular apps using a specific ad network library can exhibit serious fraud behaviours. 
For example, app \emph{Thermometer}~\cite{fraudAppThermo}, with 5-10 million downloads in Google Play, is flagged by \toolname as implementing an interaction fraud. We also investigate user reviews of some fraudulent apps and confirmed that users have submitted various complaints about their ad fraud behaviours~\cite{fraudAppFont, fraudAppThermo, fraudAppTempleqqq}. This evidence suggests that ad networks should not only publish explicit policies to regulate the usage of their ad networks, but also introduce reliable means to detect policy-violated cases for mitigating the negative impact on users and advertisers.


\vspace{-0.1in}
\subsubsection{Detection Results of VirusTotal}
We have uploaded all the detected 335 fraudulent apps to VirusTotal to explore how many of them could be flagged by existing anti-virus engines. There are 174 apps (51.9\%) labelled as \emph{AdWare} by at least one engine. Some apps are even flagged by more than 30 engines. For example, app ``com.zlqgame.yywd'' was flagged by 33 engines~\cite{adware1} and app ``com.scanpictrue.main'' was flagged by 30 engines~\cite{adware2}. However, roughly half of these fraudulent apps are not flagged, which suggests that ad fraud behaviours cannot be sufficiently identified by existing engines, especially for dynamic interaction frauds as 87.5\% (141 out of 161) of these un-flagged apps contain only such frauds.

%


%

\vspace{-0.1in}
\subsubsection{Case Studies.}
We now present real-world case studies to highlight the capability of \toolname for detecting a wide range of fraud types.
Figure~\ref{fig:casestudy} illustrates nine examples of ad-fraud apps that are found by \toolname from Google Play and third-party markets. 
\begin{enumerate}[leftmargin=*]
\item An {\em Interaction Ad} fraud was spotted in app \texttt{com.android.yatree.\\taijiaomusic} where an ad view pops up spuriously above the exit dialogue.
\item A {\em Drive-by Download Ad} fraud is found in app \texttt{com.hongap.slider} where a download starts once the ad view is clicked.
\item App \texttt{com.natsume.stone.android} implements an {\em Outside Ad} fraud where the ad view is displayed on the home screen although the app was exited. 
\item App \texttt{com.funappdev.passionatelovers.frames} includes a {\em Frequent Ad} fraud with an ad view popping up every time the main activity receives an event. 
\item App \texttt{cc.chess} is identified as showing a {\em Non-content Ad} since the ad view appears on the empty part of the screen (this may confuse users into thinking that the ad is an actual content of the host app). 
\item In app \texttt{forest.best.livewallpaper}, an {\em Ad Hidden} fraud has consisted in hiding an ad view behind the Email button.
\item  An {\em Ad Size} fraud is found in app \texttt{com.dodur.android.golf} with the ad view on the right side of the screen made too small for users to read.
\item App \texttt{com.maomao.androidcrack} places three ad views on top of a page with little content, implementing an {\em Ad Number} fraud.
\item App \texttt{com.sysapk.wifibooster} implements an {\em Ad Overlap} fraud where an ad view is placed on top of four buttons of the host app. 
\end{enumerate}

\section{Discussion}

In this work, we empirically observe that ad frauds have penetrated into official and alternative markets. All ad networks are also impacted by these fraudulent practices, exposing the mobile advertising ecosystem to various threats related to poor user experience and advertisers' losses.  
We argue that our community should invest more effort into the detection and mitigation of ad frauds towards building a trustworthy ecosystem for both advertisers and end users.
\toolname contributes to such an effort by providing the building blocks in this research direction.

The implementation of \toolname, however, carries several limitations. 

\textbf{\emph{Ad state coverage.}} 
Our experiment reveals that over 90\% of ads are displayed in either the main UI state or the exit UI state, which provide a means for \toolname to optimize its UI exploration strategy in order to achieve a balance between time efficiency and UI coverage.
However, this trade-off may cause certain ad views to be missed during UI exploration.
Fortunately, \toolname provides parameters to customize the coverage (e.g., via traversal depth) and hence to improve the soundness of the approach to reach states where the ad fraud behaviour is implemented.

\textbf{\emph{Other types of ad frauds.}} 
Although we have considered nine types of ad frauds, including five new types of \textit{dynamic interaction frauds}, which have not been explored before, our taxonomy may still be incomplete since it was built based on current policies and samples available. Other emerging types of ad frauds may have been missed.
Nevertheless, the UI transition graph built by \toolname is generic and can be reused to support the detection of potential new types of ad frauds.



\vspace{-0.1in}
\section{Related Work}
This work is mainly related to two lines of research: automated app testing and ad fraud detection.

\vspace{-0.1in}
\subsection{Automated App Testing}
Automated app testing has been widely adopted for exploring apps at runtime.
Several Android automation frameworks such as Hierarchy Viewer~\cite{HierarchyViewer}, UIAutomator~\cite{UIautomator} and Robotium~\cite{Robotium} have been developed to facilitate app testing. 
One critical step to perform automated app testing is to generate reliable test inputs~\cite{Yet}.
The most straightforward means to achieve that is to follow a random strategy (i.e., the so-called random testing), where the test inputs are generated randomly.
Indeed, various tools such as Monkey~\cite{Monkey}, Dynodroid~\cite{Dynodroid}, Intent Fuzzer~\cite{Intent} and DroidFuzzer~\cite{DroidFuzzer} have been introduced to our community following this strategy.
However, random testing is likely to generate redundant events that may lead to wastes of resources.
It is also not guaranteed that random testing will reach a good coverage of the code explored, making it not suitable for some scenarios such as ad fraud detection where certain parts of the app (e.g., ad views) are expected to be covered.

Because of the aforementioned reasons, researchers have introduced a new type of strategy, namely model-based, to generate test inputs for automatically exploring Android apps.
Indeed, many research-based tools such as GUIRipper~\cite{GUIRipper}, MobiGUITAR~\cite{MobiGUITAR}, A3E~\cite{A3E}, SwiftHand~\cite{SwiftHand}, PUMA~\cite{PUMA} have been proposed to perform model-based testing of Android apps.
These tools usually leverage finite-state machines, e.g., activities as states and events as transitions, to model the app and subsequently implement a depth-first search (DFS) or breadth-first search (BFS) strategy to explore the states for generating more effective test inputs w.r.t. code coverage.
More advanced tools such as EvoDroid~\cite{EvoDroid} and ACTEve~\cite{ACTEve} use more sophisticated techniques such as symbolic execution and evolutionary algorithms to guide the generation of test inputs aiming at reaching specific points.
However, most of them need to either instrument the system or the app,  making them hard to be directly used to detect ad frauds. DroidBot~\cite{Droidbot} is a lightweight test input generator, which is able to interact with Android apps without instrumentation. It allows users to integrate their own strategies for different scenarios. Compared with them, we have implemented a more sophisticated ad view exploration strategy for automated scalable ad fraud detection in this work.

\vspace{-0.1in}
\subsection{Ad Fraud Detection} 
Ad fraud in general is the No. 1 cybercrime counted in terms of revenues generated, ahead of tax-refund fraud~\cite{Top5Cyber}. Although ad fraud has not been substantially explored in the context of mobile advertising, it has been extensively studied in the context of web advertising. 
Many research work have been proposed to pinpoint ad frauds on the web, e.g, the detection of click frauds based on network traffic~\cite{Detectives,SLEUTH1} or search engine query logs~\cite{SBotMiner}, characterizing click frauds~\cite{Ghost,malware12,divma11} and analysing profit models~\cite{PharmaLeaks}.
We believe that these approaches can provide useful hints for researchers and practitioners in the mobile community to invent promising approaches for identifying mobile ad frauds.

Existing studies on mobile ad frauds have attempted to identify ad-fraud apps where the fraudulent behaviours can be spotted statically (the so-called \emph{static placement frauds}).
As examples, Liu \textit{et al.}~\cite{DECAF} have investigated static placement frauds on Windows Phone via analysing the layouts of apps.
Crussell \textit{et al.}~\cite{Madfraud} have developed an approach for automatically identifying click frauds.
Their approach is implemented mainly in three steps: (1) building HTTP request trees, (2) identifying ad request pages using machine learning, and (3) detecting clicks in HTTP request trees using heuristic rules. 
Unfortunately, with the evolution of ad frauds, the aforementioned approaches are incapable of identifying the latest fraudulent behaviours, e.g., they cannot be used to identify \emph{dynamic interaction fraud}. 
In this work, in addition to static placement frauds, we have introduced five new types of such frauds that have not yet been explored by our community.

\vspace{-0.1in}
\section{Conclusion}
Through an exploratory study, we characterize ad frauds by investigating policies set up by ad networks. We then build a taxonomy as a reference in the community to encourage the research line on ad fraud detections. This taxonomy comprehensively includes  four existing types of static placement frauds and five new types focusing on dynamic interaction frauds, which have not been explored in the literature. We subsequently devise and implement \toolname, a tool-supported approach for accurate and scalable detection of ad frauds in Android apps based on UI state transition graph and network traffic data.
By applying \toolname to real-world market apps, we have identified 335 ad-fraud apps covering the all considered nine types of ad frauds. Our findings suggest that ad frauds are widespread across markets and impact various markets. To the best of our knowledge, \toolname is the first attempt towards mitigating this threat to the equilibrium in the mobile ecosystem.

\bibliographystyle{ACM-Reference-Format}
\bibliography{fraudroid}

%
%

\balance
\bibliographystyle{ACM-Reference-Format}

\end{document}